# Turbulence in Molecular Clouds

JOHN DUBINSKI[1,2], RAMESH NARAYAN[1] AND T. G. PHILLIPS[3]

## ABSTRACT

We generate random Gaussian turbulent velocity fields with a Kolmogorov spectrum and use these to obtain synthetic line-of-sight velocity profiles. The profiles are found to be similar to line profiles observed in molecular clouds. We suggest methods for analysing measured line profiles to test whether they might arise from Gaussian Kolmogorov turbulence.

*Subject headings:* interstellar: molecular clouds – line profiles – turbulence – numerical methods

## 1    INTRODUCTION

Maps of the velocity fields of molecular clouds (Falgarone & Phillips 1990) present a picture which is reminiscent of turbulence seen in fluid simulations (Porter, Pouqet & Woodward 1994; Falgarone et al. 1994). Velocity variations $\Delta v$ are found to depend on scale $\lambda$ following a self-similar relation, $\Delta v \propto \lambda^\alpha$, where $\alpha$ lies between 0.3 and 0.5. A particularly simple picture of turbulence in a homogeneous, incompressible fluid was described by Kolmogorov (1941). In this picture, a fluid which is agitated by some process on a given scale sets up a cascade of turbulent vortices of diminishing size, where the characteristic velocity on scale $\lambda$ goes as $v_\lambda \sim \lambda^{1/3}$. In the language of spectral analysis, the Kolmogorov velocity field corresponds to a Gaussian random field with a steep power spectrum, $P_v(k) \equiv \langle |v_k|^2 \rangle \sim k^{-11/3}$. Modern theories of turbulence have gone much beyond Kolmogorov's ideas, with increasing emphasis being placed these days on phenomena related to "intermittency." Nevertheless, we feel that it is useful to compare the turbulence characteristics of molecular clouds to Kolmogorov's original ideas since this is still the simplest description that we have of a turbulent fluid.

In this paper, we are interested in testing the possibility that the observed velocity fields of molecular clouds follow a Kolmogorov spectrum. To do this we set up 3 dimensional random realizations of turbulent velocity fields and then synthesize observed maps of velocity profiles by projecting the data down one face of the cube assuming that the fluid is of constant density and optically thin. We use these maps of line profiles to demonstrate new observational tests for the Kolmogorov hypothesis. We also compare our results to the similar analysis by Falgarone et al (1994) of the 3D fluid simulations of compressible turbulence of Porter et al. (1994). We defer a full analysis of molecular cloud data to a future paper.

---

[1]Harvard–Smithsonian Center for Astrophysics, 60 Garden St., Cambridge, MA 02138

[2]current address: Lick Observatory, Natural Sci. II, U. of California, Santa Cruz, CA 95064

[3]Caltech, Dept. of Physics, MS 320-47, Pasadena, CA 91125

## 2   A Realization of a Turbulent Field

### 2.1   Setting up the Field

Our goal is to set up a random realization of a turbulent velocity field. The Kolmogorov power spectrum of the velocity field for homogeneous incompressible turbulence is

$$P_v(k) \equiv \langle |v_k|^2 \rangle \sim k^{-11/3}, \tag{1}$$

where $k$ is the wave number. Since the fluid is incompressible, we recognize that the flow is divergence free, $\nabla \cdot \mathbf{v} = 0$. Therefore we may define a vector potential, $\mathbf{A}$, from which we can derive the velocity field through $\mathbf{v} = \nabla \times \mathbf{A}$. The components of $\mathbf{A}$ are again described by a Gaussian random field. By dimensional arguments, the corresponding power spectrum is

$$\langle |A_k|^2 \rangle \sim k^{-17/3}. \tag{2}$$

The dispersion of $|\mathbf{A}|$ at a field point diverges for a power law this steep, so we impose a small $k$ cut-off to assure convergence. We introduce a cut-off wavenumber $k_{min}$ and redefine the vector potential power spectrum as

$$\langle |A_k|^2 \rangle = C(k^2 + k_{min}^2)^{-17/6}, \tag{3}$$

where $C$ is a constant. The scale $\lambda_{max} = 2\pi/k_{min}$ represents the "outer scale" of the turbulence, the scale on which the medium is stirred by whatever feeds the turbulence. In an interstellar molecular cloud we expect $\lambda_{max}$ to be comparable to the size of the cloud. In our simulations, we take $\lambda_{max}$ equal to the size of the computational box.

We generate random realizations of the turbulent velocity field by borrowing the methods used in numerical cosmology to set up Gaussian random density fields (e.g. Efstathiou et al. 1985). We randomly sample components of the Fourier transform of the vector potential, $\mathbf{A}_k$, in a periodic cube with equally spaced lattice points. The real and imaginary components of $\mathbf{A}_k$ at coordinate $(k_x, k_y, k_z)$ are sampled by selecting an amplitude from a Rayleigh distribution with dispersion given by the power spectrum in equation (3) and a phase angle uniformly distributed between 0 and $2\pi$. (This is equivalent to sampling from a cylindrical bivariate Gaussian.) The Fourier transform of the velocity field $\mathbf{v}_k$ is then computed by taking the curl of $\mathbf{A}$ in the Fourier domain:

$$\mathbf{v}_k = -i\mathbf{k} \times \mathbf{A}_k. \tag{4}$$

In practice, we only need the line-of-sight component of the velocity field for comparison to observations so only two components of the vector potential are generated. The velocity field in real space is finally obtained by taking the inverse transform of $\mathbf{v}_k$. We do the transform using a fast Fourier transform (FFT) inside a $256^3$ box.

### 2.2 Simulated Observations

Once we have generated the velocity field, we make synthetic velocity profiles by binning the velocities with equal weighting in $N \times N \times 256$ unit volumes across the face of the cube, as described by Falgarone et al. (1994). We have used $N = 8$ and 16 to produce $32 \times 32$ and $16 \times 16$ pixel maps (Figure 1). These velocity profiles would correspond to measured line profiles for a homogeneous cloud that is optically thin.

Since our velocity field has been generated through a Gaussian random process, the average of all the velocity profiles should be a Gaussian with dispersion,

$$\sigma_v^2 = \frac{1}{2\pi^2} \int_0^\infty k^2 dk P_v(k) = \frac{3^{1/2} C}{10\pi^{1/2}\Gamma(2/3)\Gamma(5/6)} k_{min}^{-2/3}. \tag{5}$$

For the realizations, we have used $C = 1$ and $k_{min} = 2\pi/L$ where $L$ is the boxlength set to $L = 1$. The theoretical and measured dispersions of the line-of-sight velocity agree to 1% and are $\sigma_v = 0.13$. The velocity profiles in Figure 1 are just probability distribution functions (PDFs) of the velocity along the line-of-sight. The map shows that many of the velocity profiles have strongly non-Gaussian features such as skewness, kurtosis, double peaks etc. These velocity maps closely resemble the intermediate time results of the fluid simulations of Falgarone et al., confirming that they were seeing the development of a Kolmogorov turbulent cascade in their simulations. Our results also visually resemble the maps of real molecular clouds shown in Falgarone & Phillips (1990). Because of the steepness of the Kolmogorov spectrum, our purely random realizations are able to produce quite a variety of distortions of the line profiles just as in the observations. Based on this, it would appear that the observed profiles are qualitatively similar to a Gaussian random turbulent velocity field.

Not only do the velocity profiles in Figure 1 deviate strongly from a Gaussian but there is also a spatial correlation of the profile distortions. We have quantified this effect by fitting the profiles with an expansion in Gaussian-Hermite (GH) polynomials. These functions are an orthonormal basis set which are usually applied in the context of the harmonic oscillator in quantum mechanics, but they are also handy fitting functions for our purpose. There are different definitions for the GH polynomials but we use the definition prescribed by Van der Marel & Franx (1993) in their analysis of the line profiles in galaxy spectra. The line profile is first fit to a Gaussian to determine the best fit of the mean velocity, $V$, and the dispersion, $\sigma$. With these parameters fixed, the profile is then fit with a higher order expansion,

$$f(v) = \frac{\gamma\alpha(w)}{\sigma}\left\{1 + \sum_{j=3}^{N} h_j H_j(w)\right\}, \quad w = \frac{(v - V)}{\sigma}, \tag{6}$$

where $\gamma$ is a normalizing constant, $\alpha(w) = (2\pi\sigma^2)^{-1/2}\exp(-w^2/2)$ is the best-fitting Gaussian, and $H_j(w)$ are the GH polynomials. The additional parameters we wish to fit are the coefficients $h_j$. The coefficients $h_1$ and $h_2$ are set to zero because the process of first fitting the Gaussian $\alpha(w)$ to the individual profiles includes their effect on the function. In general, the even coefficients broaden or narrow the profiles while the odd coefficients add skew in either direction depending on the sign. The coefficients $h_3$ and $h_4$ are similar to the third and fourth moments of the distribution which Falgarone et al. used to characterize the data. Figure 2 shows a subsample of the profiles in Fig. 1 fitted up to the fourth order (i.e. including $h_3$ and $h_4$). We see that the fits are fairly good. The GH expansion is therefore a useful way to characterize the shapes of the line profiles.

fig-map16.ps

FIG. 1.–Simulated $16 \times 16$ map of velocity profiles created by binning velocities in $16 \times 16 \times 256$ rods on the face of a $256^3$ cube. Although the mean of all these profiles is Gaussian, the individual profiles show noticeable deviations from the mean Gaussian.

TABLE 1
MAP COEFFICIENTS - $\alpha = 0.33$

| Map | $V$ | | $\sigma$ | | $h_3$ | | $h_4$ | |
|-----|------|------|------|------|------|------|------|------|
|     | mean | rms | mean | rms | mean | rms | mean | rms |
| $16 \times 16$ | 0.000 | 0.062 | 0.111 | 0.021 | 0.001 | 0.054 | -0.008 | 0.033 |
| $32 \times 32$ | 0.000 | 0.070 | 0.121 | 0.024 | 0.001 | 0.077 | 0.011 | 0.053 |

Figure 3 shows maps of $V$, $\sigma$, $h_3$ and $h_4$ for one of our random velocity fields. The velocities are binned in rods of dimension $8 \times 8 \times 256$ elements across the face of the cube to produce $32 \times 32$ velocity profile maps. Maps at $16 \times 16$ resolution show the same features. The map of $V$ immediately shows the effect of imposing a small wave number cut-off, $k_{min}$. For this realization, we selected $k_{min}$ equal to the maximum wavelength of the box i.e. $k_{min} = 2\pi/L$, where $L$ is the box length. The steepness of the spectrum suggests that the velocity field should be described by a dominant wave of this length with superimposed small scale ripples. The $V$ map does show a dominant wave. The portions of the map with mean velocities moving away and towards us are spatially connected revealing the dominance of one large scale as imposed by $k_{min}$. Indeed the net impression is that the fluid is rotating around an axis pointed towards the northeast. This makes one wonder whether some of the cloud rotations that have been claimed are merely random turbulent eddies on the outer scale of the cloud. The map of $\sigma$ shows that the regions with broader or narrower than average peaks tend to be spatially clustered in different regions of the map. The maps of $h_3$ and $h_4$ are much noisier and do not show much spatial correlation. These higher order effects can apparently vary significantly from point to point.

Another useful statistic is the map average and rms values of $V$, $\sigma$ and the coefficients $h_3$ and $h_4$ for an ensemble of realizations. This should give a quantitative feel for the degree of non-Gaussian deviations to be expected in a map and may be a useful statistic for comparison to real data and fluid simulations. The average of the mean and rms values of the coefficients for an ensemble of 10 maps are given in Table 1. The mean velocity is zero as expected and the rms of the mean velocity is about half the 3D value of the expected dispersion, $\sigma_v = 0.13$. The mean value of $\sigma$ for the projected profiles is approximately equal to the 3D value. The $h_3$ and $h_4$ map values suggest that there is an approximately 5% to 8% deviation from the best-fitting Gaussian. The $16 \times 16$ map has slightly smaller values for all of the coefficients showing the effect of a large effective smoothing radius.

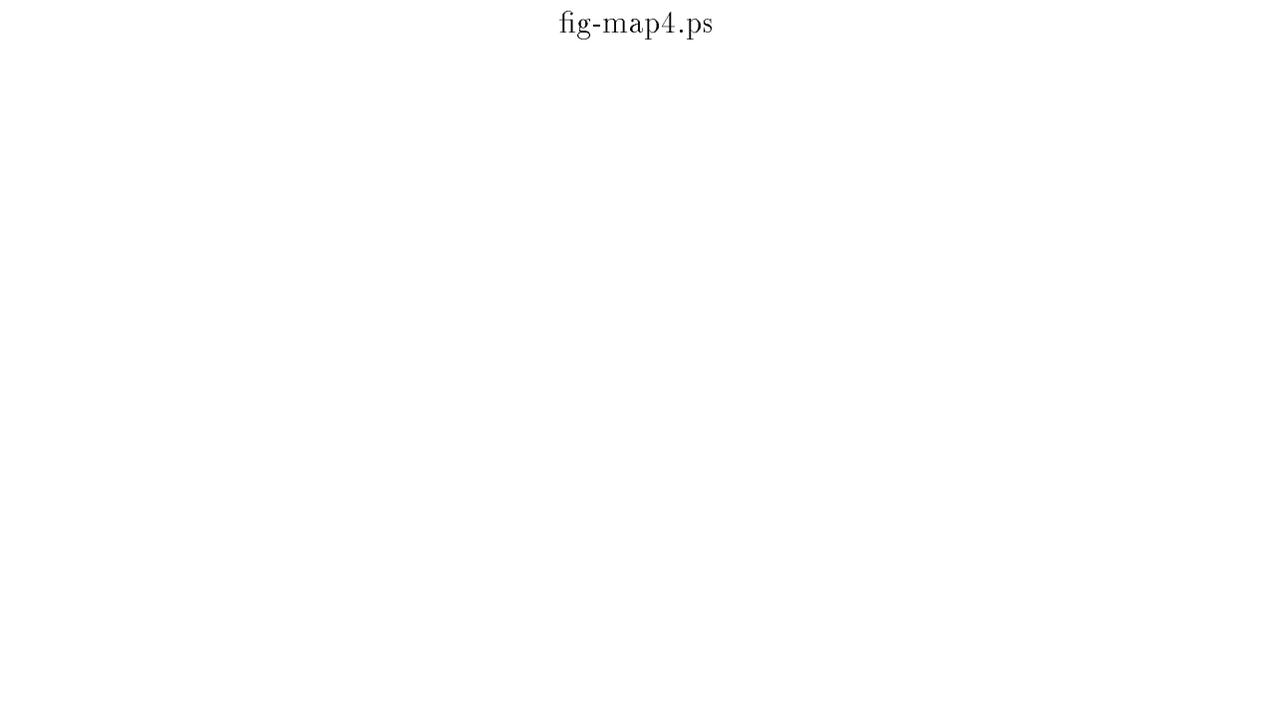

Fig. 2.–A blow-up of the profiles (*solid*) in the lower right corner of Fig. 1 with GH expansion fits to 4th order (*dashed*). The fits are generally good (except for the double–humped profile at the bottom), demonstrating that this technique is a good way to quantify the non-Gaussian deviations of the line profiles. Higher order expansions can improve the fits but the the statistics of these fits should be interpreted with caution.

fig-contmap.ps

Fig. 3.–Contour maps of $V$, $\sigma$, $h_3$, and $h_4$ for a Kolmogorov turbulent velocity field. The contours in $V$ extend between -0.20 and 0.20 in steps of .02, $\sigma$ between 0 and 0.30 in steps of 0.02, $h_3$ between -0.5 and 0.5 in steps of 0.05 and $h_4$ between -0.25 and 0.25 in steps of .025. The $V$ map shows a dominant wave of length $\lambda_{max} = 2\pi/k_{min}$, as expected for this steep power spectrum, with regions approaching and receding being spatially connected. The $\sigma$ map shows peaks and valleys corresponding to the broadest and narrowest distributions in Fig. 1. The $h_3$ and $h_4$ maps show much less spatial correlation on large scales than $V$ or $\sigma$.

A comparison of the maps of $V$ and $h_3$ can also be used to discriminate between clouds that are turbulent and those that are rotating. For a rotating cloud, velocity profiles are asymmetric because of an extended low-velocity tail from gas observed in front of and behind the tangent point along the line-of-sight. One therefore expects $V$ and $h_3$ to have opposite signs from point to point in a rotating cloud; the maps of $V$ and $h_3$ should be similar and should resemble each other except for a sign change. Figure 3 shows that when there is turbulence the field of $h_3$ is incoherent and does not resemble the more regular $V$ field. This diagnostic can in principle be used to distinguish a turbulent medium from one with a dominant net rotation. (We thank the referee, R. van der Marel, for pointing this out.)

Fits with higher order coefficients, $h_i$, are also possible and may contain additional information. We have made fits with coefficients up to $h_8$ and have succeeded in fitting the double humped profiles in Figure 2. The fitted values of $h_3$ and $h_4$ do not change significantly when including these higher order terms so the results of Table I remain unchanged. However, care should be taken in interpreting the maps with $h_5$ and higher. A simple mean and rms of the map can be misleading, since the distribution of higher order $h_i$ across the map can be highly non-Gaussian. In most cases where the profiles are featureless and symmetric, the values of higher order $h_i$ quickly die off. Profiles with many bumps and wiggles tend to have large amplitude $h_i$ which alternate in sign between even and odd coefficients as one goes to higher order. The higher order coefficients are therefore not very informative. Fits to $h_4$ and perhaps as high as $h_8$ are all that is required to quantify the non-Gaussian behavior of the velocity profiles.

### 2.3 A Steeper Spectrum

Observations of real molecular clouds suggest velocity scaling relations $v \propto \lambda^\alpha$ with $\alpha$ as large as 0.5 (Falgarone & Phillips 1990). For $\alpha = 0.5$, the velocity power spectrum is steeper, going as $P_v(k) \sim k^{-4}$. This steeper spectrum should show more non-Gaussian features in the line-of-sight velocity profiles than we have seen with the Kolmogorov spectrum. We repeated the previous calculations using a $k^{-4}$ velocity spectrum, again introducing a vector potential with cut-off at $k_{min}$ as in equation 3. In this case, the rms variation in the velocity is (cf. equation 5)

$$\sigma_v^2 = \frac{C}{8\pi} k_{min}^{-1}. \tag{7}$$

With C=1 and $k_{min} = 2\pi$, $\sigma_v = 0.079$. Table 2 shows the values of mean and rms map coefficients for the steeper spectrum for comparison to Table 1. The rms value of $V$ is larger relative to $\sigma$ than in the previous case. The rms values of $h_3$ and $h_4$ are also larger, indicating more non-Gaussian features in the line profiles than expected for the Kolmogorov spectrum.

### 3  COMPARISON WITH SIMULATIONS OF COMPRESSIBLE TURBULENCE

Falgarone et al. (1994) have recently analyzed Porter et al.'s (1994) three-dimensional fluid simulations of compressible turbulence with a moderate Mach number (initial rms value of 1.1) to compare the resulting simulated velocity field to the observations of velocity profiles in molecular clouds. They inject energy into the fluid modelled as a perfect gas with $\gamma = 4/3$ at $t = 0$ and let it evolve freely, following the development of shocks using the Piecewise-Parabolic method. A turbulent velocity field quickly develops and energy is dissipated in the shocks. At various times, they produce simulated line-of-sight observations of their computational box for comparision to molecular cloud observations.

TABLE 2
MAP COEFFICIENTS $- \alpha = 0.5$

| Map | $V$ | | $\sigma$ | | $h_3$ | | $h_4$ | |
|---|---|---|---|---|---|---|---|---|
| | mean | rms | mean | rms | mean | rms | mean | rms |
| $16 \times 16$ | 0.000 | 0.043 | 0.069 | 0.013 | 0.000 | 0.084 | -0.016 | 0.059 |
| $32 \times 32$ | 0.000 | 0.045 | 0.068 | 0.015 | 0.001 | 0.107 | -0.017 | 0.081 |

The velocity profiles are initially non-Gaussian as the gas begins to dissipate after its initial agitation. As the gas settles down, the non-Gaussian features gradually disappear. The intermediate snapshot in their simulation produces velocity profiles very similar to the observations and the power spectrum is dominated by energy in incompressible modes closely following a Kolmogorov scaling, $E(k) \propto k^{-5/3}$. The close resemblance of their measured velocity profiles to those in Figure 1 further supports the idea that their simulations of turbulence are settling to the Kolmogorov spectrum.

Falgarone et al. (1994) also pointed out that there was evidence for the process of intermittency in their simulations. They measured a non-Gaussian tail in the distribution of vorticity, $\omega = |\nabla \times \mathbf{v}|$, as measured along the line of sight as well as in phase space plots of the vorticity versus the 3D bulk velocity at different points along the line of sight. In particular, the appearance of dense striations extending to large vorticity for fixed bulk velocity suggested the localized process of intermittency. They noted that this behavior could also be responsible for some of the non-Gaussian features of the velocity profiles though the large variations in bulk flow along the line of sight due to steepness of the Kolmogorov spectrum were the dominant effect in producing the profiles.

We do not expect to see streaks in the $\omega - v$ phase space with a Kolmogorov spectrum since the underlying random process generating the field is Gaussian (though "non-Gaussian" velocity profiles are created because of the steep power spectrum). Figure 4 shows a plot of the vorticity, $\omega$, versus the three components of velocity along 3 adjacent lines of sight down the $x$ face of the cube. They are chosen because of their pronounced non-Gaussian velocity profiles. Also, these synthesized profiles were generated at $128^3$ resolution because the extra memory overhead to generate, $\omega$, at $256^3$ resolution was too great for our local resources. These plots are similar to Figure 9 in Falgarone et al. (1994). The Kolmogorov turbulent fields show no signs of the non-Gaussian streaks in $\omega - v$ phase space that are seen in the fluid simulations.

Figure 5 emphasizes again why we expect non-Gaussian profiles from an underlying

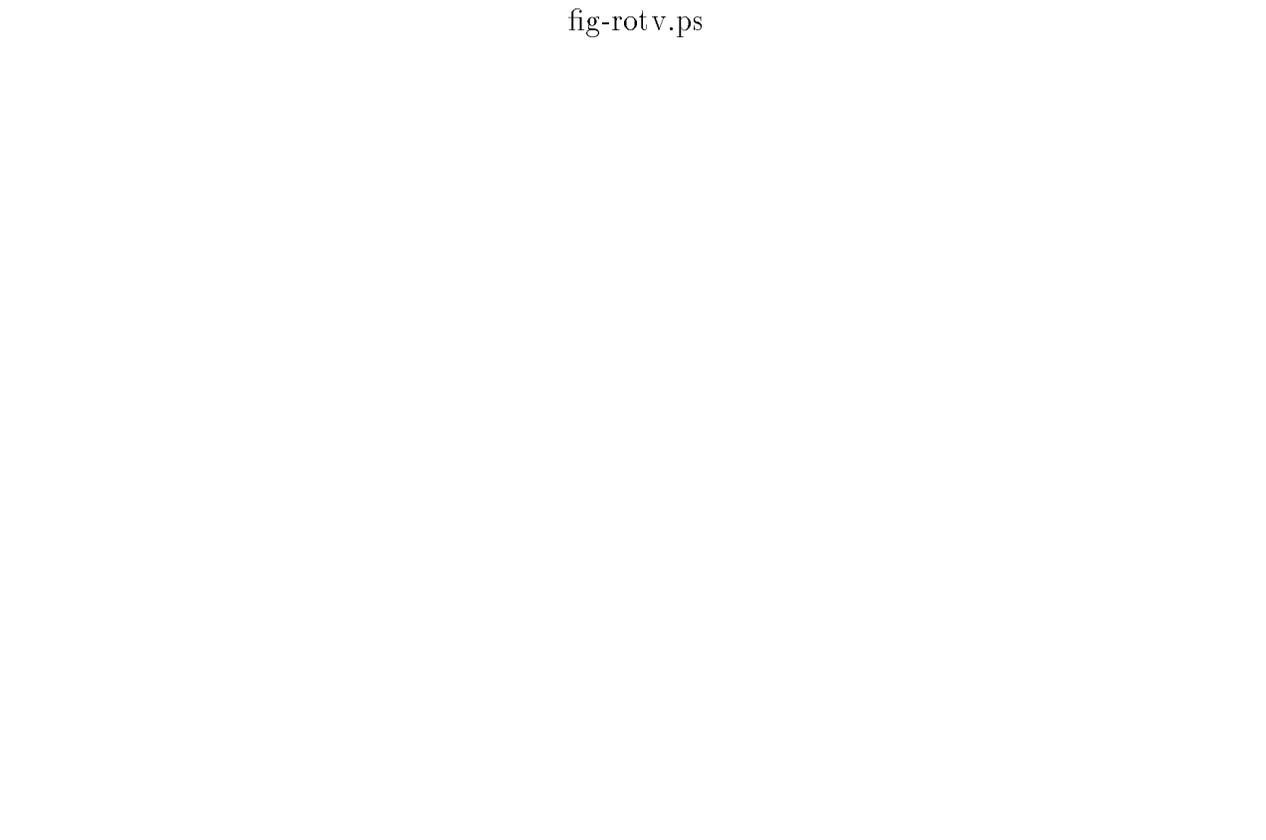

Fig. 4.–Scatter plot of |rot $v$| vs. the velocity components $v_x$, $v_y$, and $v_z$ along 3 adjacent lines of sight averaged including $8 \times 8$ pixels on the face of $128^3$ random velocity field. Also shown are the line-of-sight velocity profiles of the $v_x$ component. There are no signs of the striations in phase space seen by Falgarone et al. 1994 in their fluid simulations which they suggested were signs of the non-Gaussian effects of intermittency. The purely Gaussian turbulent field has a bland appearance in this representation of phase space.

Gaussian process. This figure shows the distribution of the line-of-sight component of velocities in the $8 \times 8$ pixel region for the three profiles in Figure 4. The large scale wave of the bulk flow imprinted by the dominant long waves of the Kolmogorov spectrum is once again demonstrated.

## 4   CONCLUSIONS

Despite the simplifying assumptions of homogeneity, isotropy and incompressibility, Kolmogorov turbulence seems to be a good description of the velocity profiles of molecular clouds. From our analysis, we draw the following conclusions:

1. The line-of-sight velocity profiles derived from an optically thin Kolmogorov turbulent velocity field exhibit non-Gaussian features which resemble the features seen in real molecular clouds and simulations of compressible turbulence. These features occur because of the steepness of the Kolmogorov power spectrum i.e. a few large scale waves dominate the character of the bulk flow.

2. The deviation from a Gaussian of the velocity profiles as quantified by higher order fits with Gaussian-Hermite polynomials are at the 5% to 10% level (rms value of the $h_3$ and $h_4$ coefficients over the simulated map). A steeper velocity spectrum ($P_v \sim k^{-4}$) creates more non-Gaussian features in the line profiles which show up as larger values of $h_3$ and $h_4$.

3. The presence of an outer scale, $\lambda_{max}$, to the turbulent field reveals itself as a large scale wave in the mean projected velocity, mimicing a net rotation. A comparison of the $V$ and $h_3$ fields derived from the line-of-sight profiles can be used to discriminate between rotation and turbulence in the cloud: a strong anti-correlation between $V$ and $h_3$ indicates that rotation dominates, while a regular $V$ field with an incoherent $h_3$ field indicates that the apparent rotation is merely part of a turbulent cascade.

4. Fluid simulations show evidence of intermittency as revealed by non-Gaussian stripes in plots of vorticity versus bulk velocity. These features do not occur in our simple model of Kolmogorov turbulence because it is a fundamentally Gaussian model. Given that our model produces line profiles that are fairly similar to actual line profiles of molecular clouds, it is not clear if the data provide evidence for intermittency.

We thank the referee R. van der Marel for useful comments and suggestions. TP and RN acknowledge NSF grants AST 93-13929 and AST 9148279 respectively. JD recognizes a CfA postdoctoral fellowship.

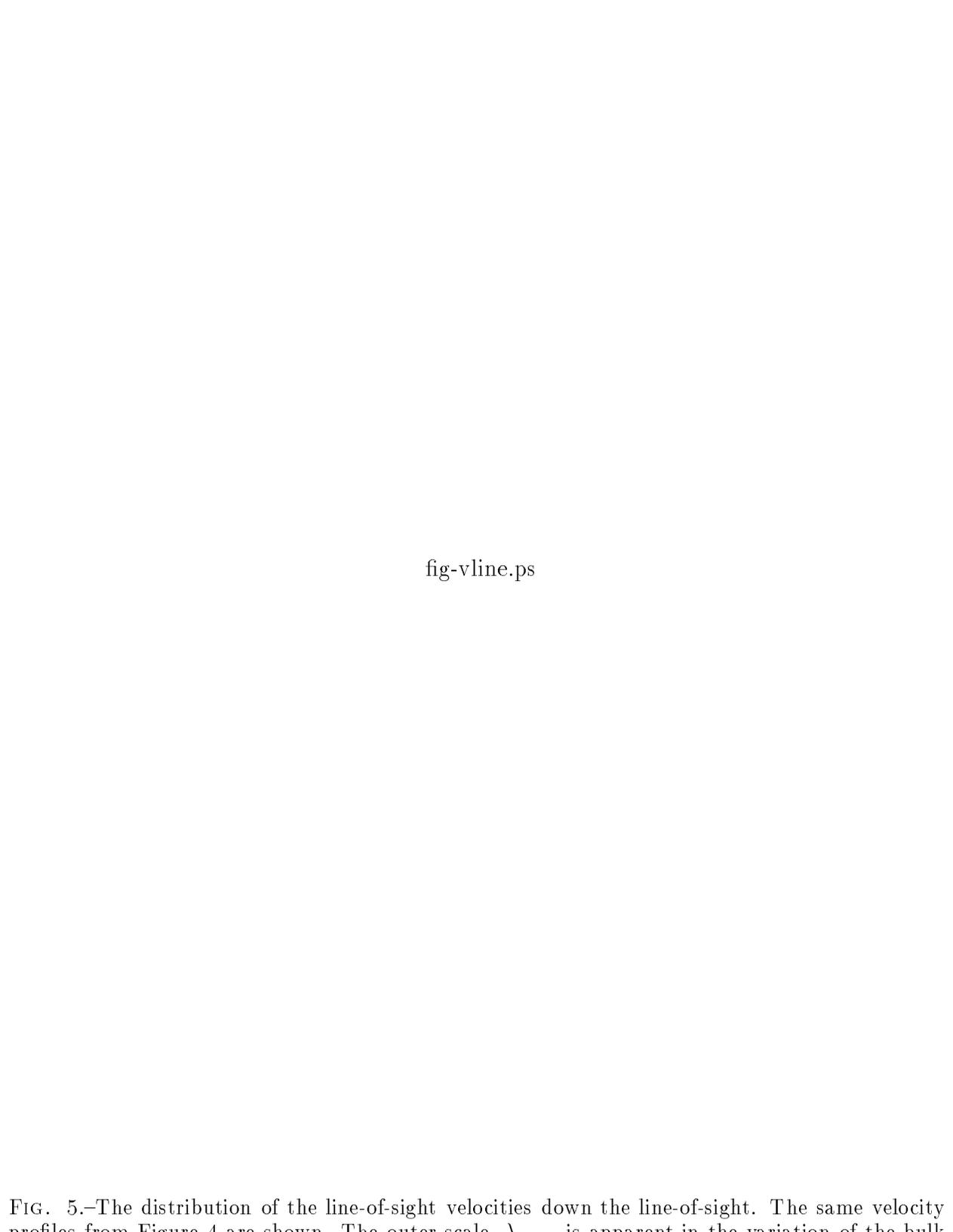

Fig. 5.–The distribution of the line-of-sight velocities down the line-of-sight. The same velocity profiles from Figure 4 are shown. The outer scale, $\lambda_{max}$, is apparent in the variation of the bulk velocity down the line of sight and is responsible for the observed non-Gaussian velocity profile.